# VaxPulse: Monitoring of Online Public Concerns to Enhance Post-licensure Vaccine Surveillance

Muhammad JAVED[a,b,1], Sedigh KHADEMI[a,b], Joanne HICKMAN[a,b], Jim BUTTERY[a,b,c], Hazel CLOTHIER[a,b], Gerardo Luis DIMAGUILA[a,b,c]

[a]*Epidemiology Informatics, Centre for Health Analytics, Melbourne Children's Campus, Parkville, Victoria, Australia*
[b]*Surveillance of Adverse Events Following Vaccination In the Community (SAEFVIC), Murdoch Children's Research Institute, Parkville, Victoria, Australia*
[c]*Department of Paediatrics, The University of Melbourne, Parkville, Australia*

ORCiD ID: Muhammad Javed https://orcid.org/0000-0002-7022-6596, Sedigh Khademi https://orcid.org/0000-0001-6146-1415, Joanne Hickman https://orcid.org/0000-0002-1862-1708, Jim Buttery https://orcid.org/0000-0001-9905-2035, Hazel Clothier https://orcid.org/0000-0001-7594-0361, Gerardo Luis Dimaguila https://orcid.org/0000-0002-3498-6256

**Abstract.** The recent vaccine-related infodemic has amplified public concerns, highlighting the need for proactive misinformation management. We describe how we enhanced the reporting surveillance system of Victoria's vaccine safety service, SAEFVIC, through the incorporation of new information sources for public sentiment analysis, topics of discussion, and hesitancies about vaccinations online. Using VaxPulse, a multi-step framework, we integrate adverse events following immunisation (AEFI) with sentiment analysis, demonstrating the importance of contextualising public concerns. Additionally, we emphasise the need to address non-English languages to stratify concerns across ethno-lingual communities, providing valuable insights for vaccine uptake strategies and combating mis/disinformation. The framework is applied to real-world examples and a case study on women's vaccine hesitancy, showcasing its benefits and adaptability by identifying public opinion from online media.

## 1. Introduction

SAEFVIC (Surveillance of Adverse Events Following Vaccination in the Community) is the jurisdictional vaccine safety system of Victoria, Australia [1] that facilitates spontaneous reporting, assessment, and analysis of adverse events following immunisation (AEFI) trends. SAEFVIC provides education for the public and immunisers through eLearning, seminars, online resources, and updated immunisation guidelines. To ensure timely vaccine safety information, the Victorian vaccine safety report was developed and updated weekly with rates of AEFI and related resources.

---

[1] Corresponding Author: Muhammad Javed, Muhammad.javed@mcri.edu.au

WHO identified vaccine hesitancy as one of the top 10 global health threats [2], often fuelled by concerns about vaccine safety, skepticism of science, and social media disinformation. Social media's accessibility has made it a significant platform for sharing views and a vital source of public health information. Most vaccine-related information on social media is not from official sources, spreading mis/disinformation that fuels hesitancy and the vaccine infodemic. Moreover, people's attitudes towards immunisation are dynamic and require constant monitoring to guide effective communication strategies [3]. Adverse events following immunisation (AEFI) and other vaccine hesitancy factors remain prominent discussion topics on social media [4].

While AEFI is regarded as a major contributor to hesitancy [5], only a few studies have processed AEFI-related data. Moreover, they do not analyse sentiments based on vaccine AEFI. To address this gap, we developed an AEFI-centric machine learning framework to generate valuable insights for vaccine surveillance groups and infodemic management. We describe our process to identify, collect, and analyse AEFI-based public sentiments, i.e., concerns and hesitancies, about vaccines from various online data sources, to generate actionable insights. These insights will support our initiatives in managing the vaccine infodemic and provide vaccine confidence and uptake groups in Victoria and other jurisdictions in Australia a near real-time source of information on current public concerns about vaccines.

We evaluated sentiment analysis and topic modelling methods, focusing on the use of BERT transformer-based models and Large Language Models (LLMs). These approaches were chosen for their ability to enhance the identification of social media sentiments and topics of discussion.

Common sentiment analysis techniques, like those using TextBlob and Vader, often struggle with context-specific language. Supervised machine learning models, such as Naïve Bayes, can be limited by data sparsity. To analyse social media discussions about vaccines accurately, it's crucial to consider the dynamic nature of the topic and the specific vaccine. Categorising comments by vaccine can better identify relevant sentiments and hesitancies. Fine-tuning pre-trained language models like BERT is a common NLP technique for tasks such as classification [6]. BERT has demonstrated superior performance compared to traditional machine learning models [7].

## 2. Methods

Topic modelling identifies abstract topics within documents. Recent advancements, like BERTopic [8], leverage BERT and transformer embeddings to create dense topic clusters, incorporating context-specific word meanings. BERTopic's latest version allows fine-tuning topic representations using LLMs.

Existing research focuses on general public opinion. To understand vaccine-specific concerns, we need to identify AEFI-related topics. Topic modelling can categorise social media posts by AEFI theme. These AEFIs have had a substantial impact on vaccine hesitancy and policy decisions in the past.

To effectively address online vaccine concerns, sentiment analysis should consider the specific AEFI context to guide appropriate responses. Our proposed method adds these key steps, illustrated in Figure 1.

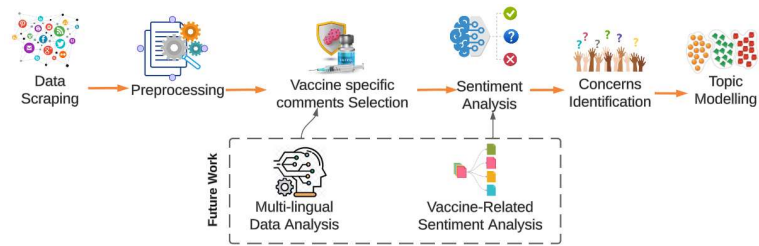

**Figure 1.** VaxPulse framework

To identify diverse community concerns, we are currently extracting data from X, Reddit, YouTube, Facebook, and Google Trends using authenticated APIs. We leverage Tweepy for X, PRAW for Reddit, and Google API Client Library for YouTube.

Language ambiguity can hinder accurate search results. To overcome this, we employ a two-step approach. First, we categorise comments using an ensemble technique (fine-tuned Roberta-large-mnli and GPT-4o models) into general vaccine discussions, personal experiences, or unrelated topics. Second, we utilise LLMs like GPT-4o with prompt engineering to identify specific vaccine-related comments.

We fine-tuned CT-BERT V2 to classify social media comments about vaccine AEFIs into negative, neutral, and positive, achieving 93% accuracy on 13,715 annotated posts. To further improve performance, we combined this model with GPT-4o for comments with low BERT confidence, significantly enhancing overall accuracy. We leverage LLMs to identify and categorise public vaccine concerns expressed online and apply BERTopic to identify topics from grouped comments under each concern identified. Negative and neutral comments are also analysed separately to capture hesitations and information-seeking behaviours.

## 3. Results

This section presents results from applying our method to analyse vaccine hesitancy expressed in online and social media posts related to AEFI. While we initially focused on COVID-19 vaccine concerns, we have since expanded our analysis to other vaccines Shingrix®, RSV, and HPV vaccines.

We retrieved social media comments (n= 871,596) about COVID-19, Shingrix®, and RSV vaccines' adverse effects posted between August 2023 and July 2024. After cleaning the data (removing links, mentions, special characters, and short sentences), we analysed 819,388 comments to understand current online trends. Despite employing distinct data extraction pipelines for each vaccine, using both an ensemble technique and LLMs allowed us to segregate 705,145 COVID-19 vaccine-related comments. We found 9.87% negative, 27.23% positive, and 62.90% neutral sentiments toward COVID-19 vaccines. We noted that this is a shift from what we discovered previously [9], from more negative to more neutral sentiments, suggesting increased public interest in vaccine information and potential adverse effects. We identify and categorise public vaccine concerns. Key concerns include safety, side effects, misinformation, trust in authorities, and previous negative experiences. We found that these concerns persist; mis/disinformation remains a major issue, especially with co-administered vaccines.

Table 1 presents a sample of discussions related to serious AEFIs, while Table 2 contains some hesitancies identified among female individuals.

**Table 1.** Sample topics related to serious AEFIs from the aggregate dataset.

| S.No | Topic of Discussion | Comments Count |
| --- | --- | --- |
| 1 | Vaccine-Related Deaths | 1907 |
| 2 | Blood Clot Concerns and Experiences | 1844 |
| 3 | Myocarditis and Pericarditis Following COVID-19 Vaccination | 2574 |

**Table 2.** Sample hesitancies that could pertain to the female sex for COVID-19 vaccines

| S.No | Topic of Discussion | Comments Count |
| --- | --- | --- |
| 1 | Irregular Menstrual Cycles | 5023 |
| 2 | Experiences and Recommendations for Vaccines During Pregnancy | 569 |
| 3 | COVID Vaccine and Miscarriage Concerns | 67 |

A notable use of VaxPulse's insights is with regards to AEFIs concerning female sex, with the most common being menstrual cycle abnormalities, illustrated in Figure 2. These peaked around November 2023 (below 800 comments) and May 2024 (approximately 1200 comments). SAEFVIC investigated menstrual changes post COVID-19 vaccination, and found there was an association. As reports revealed that individuals who menstruate felt dismissed and distressed due to clinicians' scepticism about a potential link to vaccination [10], SAEFVIC developed TikTok and Instagram videos to reassure women of their concerns, and to educate that the effects were only temporary [11]. This highlights the need for continuous social media monitoring and targeted communication strategies to build public trust in vaccination.

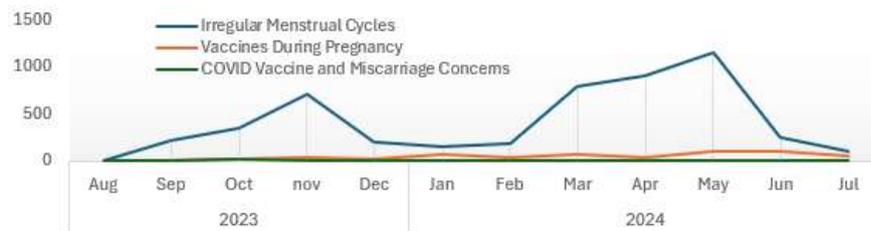

**Figure 2.** Female-related hesitancy topics over time

In addition, we also identified that social bots influenced COVID-19 vaccine discussions [9]. We found that bots contributed 23.72% of tweets and amplified AEFIs in human posts.

## 4. Discussion

Online media enables rapid exchange of health information and facilitates public engagement. Online and social media discourse analysis is a cost-effective tool for gathering population-level data, complementing traditional methods. Social media significantly influences public health information and vaccine decisions. However, it also contributes to the spread of misinformation and negative sentiment regarding vaccine adverse effects, potentially fuelling vaccine hesitancy. Our AEFI-centric machine learning framework allows for targeted identification of vaccine concerns.

Long-term trend analysis can reveal the underlying causes of vaccine-related concerns in specific communities. This will help provide timely, accurate, and relevant information to the public, fostering informed decision-making and reducing vaccine anxiety. In the future, we aim to refine hesitancy types beyond simple polarity, enabling near real-time observation of public vaccine sentiment. We will use a multi-model approach to translate multilingual social media data to understand ethno-lingual community concerns, aiding both vaccine uptake researchers and misinformation combat efforts.

## 5. Conclusions

VaxPulse' AEFI-centric machine learning framework enables real-time insights to inform communication strategies, as demonstrated by SAEFVIC's investigation of menstrual changes post-COVID-19 vaccination and our detection of social bot influence on online discussions. This underscores the importance of robust social media monitoring to address persistent public concerns about vaccine safety, side effects, and trust in authorities.

## References

1. Clothier HJ, Crawford NW, Russell M, et al. Evaluation of 'SAEFVIC', a pharmacovigilance surveillance scheme for the spontaneous reporting of adverse events following immunisation in Victoria, Australia. Drug Saf. 2017;40:483-95.
2. Thangaraju P, Venkatesan S. WHO Ten threats to global health in 2019: antimicrobial resistance. Cukurova Med J. 2019;44:1150-1.
3. MacDonald NE, Dube E. Vaccine safety concerns: Should we be changing the way we support immunization? EClinicalMedicine. 2020;23:100413.
4. Hussain Z, Sheikh Z, Tahir A, Dashtipour K, Gogate M, Sheikh A, Hussain A. Artificial intelligence-enabled social media analysis for pharmacovigilance of COVID-19 vaccinations in the United Kingdom: observational study. JMIR Public Health Surveill. 2022 May 27;8(5):e32543. doi: 10.2196/32543.
5. Melton CA, Olusanya OA, Ammar N, et al. Public sentiment analysis and topic modeling regarding COVID-19 vaccines on the Reddit social media platform: A call to action for strengthening vaccine confidence. J Infect Public Health. 2021;14(10):1505-12.
6. Kovács L, Voronkov A. First-order theorem proving and Vampire. In: International Conference on Computer Aided Verification. Berlin, Heidelberg: Springer; 2013. p. 1-35.
7. Cotfas LA, Delcea C, Roxin I, et al. The longest month: analyzing COVID-19 vaccination opinions dynamics from tweets in the month following the first vaccine announcement. IEEE Access. 2021;9:33203-23.
8. Grootendorst M. BERTopic: Neural topic modeling with a class-based TF-IDF procedure. arXiv Preprint arXiv:2203.05794. 2022.
9. Javed M, Dimaguila GL, Habibabadi SK, Palmer C, Buttery J. Learning from machines? Social bots influence on COVID-19 vaccination-related discussions: 2021 in review. In: Proceedings of the 2023 Australasian Computer Science Week; 2023. p. 190-7.
10. Liew TM, Lee CS. Examining the utility of social media in COVID-19 vaccination: unsupervised learning of 672,133 Twitter posts. JMIR Public Health Surveill. 2021;7(11):e29789.
11. Social media health talks "Does COVID infection and the vaccine affect periods? Prof Sonia Grover tells us more!" [Internet]. 2024 Dec 4 [cited 2025 Mar 31]. Available from: https://www.instagram.com/reel/CzKo83mSxNg/?igsh=MXRveGdxdjE1cmEwOQ%3D%3D